\documentclass[%
 reprint,
 amsmath,amssymb,
 aps,
prm
]{revtex4-2}

\usepackage{graphicx}
\usepackage{dcolumn}
\usepackage{bm}
\usepackage{siunitx}
\usepackage{subcaption}
\usepackage{hyperref}
\usepackage{xcolor}
\usepackage{amssymb}

\begin{document}

\preprint{APS/123-QED}

\title{Weak non-linearities of amorphous polymer under creep}

\author{Martin Roman-Faure}
\author{H\'el\`ene Montes}%
\author{François Lequeux}
\author{Antoine Chateauminois}%
\email{antoine.chateauminois@espci.fr}
\affiliation{%
Soft Matter Science and Engineering (SIMM), ESPCI Paris, PSL University, Sorbonne Universit\'e, CNRS, F-75005 Paris, France
}%

\date{\today}

\begin{abstract}
The creep behavior of an amorphous poly(etherimide) (PEI) polymer is investigated in the vicinity of its glass transition in a weakly non linear regime where the acceleration of the creep response is driven by local configurational rearrangements. From the time shifts of the creep compliance curves under increasing applied stresses in the range 1-15~\si{\mega\pascal}, we determine a macroscopic acceleration factor. At the start of creep, the stress is homogeneous and the macroscopic acceleration can be assimilated to that of the local rearrangements which is shown to vary as $f=e^{-(\sigma/Y)^n} $ with $n=2 \pm 0.2$, where $\sigma$ is the local stress and $Y$ is a decreasing function of compliance. This experimental result is in agreement with the recent theory of Long \textit{et al.} (\textit{Phys. Rev. Mat.} (2018) \textbf{2}, 105601 ) which predicts $n=2$. From a mean field approximation, we interpret the variation of $Y$ with compliance as the result of the development of stress heterogneities during creep.
\end{abstract}

\maketitle
\section{Introduction}
Amorphous polymer and elastomer materials have linear mechanical properties which vary drastically around their glass transition~\cite{ferry_viscoelastic_1980}. At low temperatures, they exhibit an elastic modulus of the order of 1 GPa as a result of Van der Waal interactions between monomers~\cite{long_heterogeneous_2001}.
At high temperatures, their elastic modulus drops by three orders of magnitude towards about 1 MPa. In this domain, the elastic modulus is conferred by the entropy of the chains between two cross-linking points for elastomers~\cite{rubinstein_elasticity_2002,rubinstein_polymer_2014}, or a so-called plateau modulus between entanglements for non-cross-linked polymers~\cite{ferry_viscoelastic_1980}. Similarly, mechanical properties at high strain also vary significantly with temperature. At high temperature, in the rubbery state, the macroscopic mechanical response still arise from chain entropy. As a consequence the macroscopic mechanical response remains approximately linear over a strain amplitude of 100\%~\cite{rubinstein_elasticity_2002}. On the other hand, in the glassy state, the mechanical response originates from molecular and Van der Waal interactions between close packed monomers. Then, non-linearity find its origin in local rearrangements~\cite{argon_theory_1973,langer_shear-transformation-zone_2008,dequidt_heterogeneous_2016}. As a result, a macroscopic deformation of a few percent is enough to induce monomer rearrangements and to make the macroscopic response non-linear.\\
The transition from rubber-like to glassy regime is progressive because amorphous polymers are highly disordered systems. They are made up of nanometric domains, each of them with a specific relaxation time under thermal activation~\cite{ediger_spatially_2000}. These times are distributed over 4 orders of magnitude. Consequently, in the vicinity of the glass transition, the measurement time sweeps the distribution of local relaxation times which allows to reveal the respective role of domains with various relaxation times.\\
Closed-packed monomers may encounter local configurational changes under the effect of thermal agitation within a characteristic time $\tau$. However, under the action of a stress $\sigma$, the time required for configurational changes is accelerated~\cite{lee_deformation-induced_2009,lee_direct_2009,lee_molecular_2009,bennin_rejuvenation_2020,loo_chain_2000,perez-aparicio_dielectric_2016,perez-aparicio_dielectric_2016-1,kalfus_probing_2012,bending_measurement_2014,hebert_reversing_2017}, and becomes $f(\sigma)\tau$ where $f$ is a stress-dependent acceleration function, always smaller than $1$,  which tends towards one when $\sigma$ tends towards zero.  According to Eyring's classical model~\cite{eyring_viscosity_1936}, the acceleration function writes $f=\exp^{-\sigma/Y}$, with $Y=kT/\textit{v}$, where $kT$ is the thermal energy and $\textit{v}$ is an activation volume. However, a more recent theory~\cite{dequidt_heterogeneous_2016,dequidt_mechanical_2012,long_dynamics_2018,conca_acceleration_2017} corroborated by local optical and mechanical measurements of segmental dynamics~\cite{lee_molecular_2009} and strain relaxation experiments~\cite{belguise_weak_2021} suggests that the acceleration function is rather $f=\exp^{-(\sigma/Y)^2}$. Indeed, Eyring's expression~\cite{eyring_viscosity_1936} originates from a simple linear expansion with respect to stress of the value of the energy barrier  between two stable configurations of the monomers. In the Long~\textit{et al.} approach, the expression of the acceleration function is obtained using a different approach~\cite{long_dynamics_2018,dequidt_heterogeneous_2016,dequidt_mechanical_2012,conca_acceleration_2017} which is based on an estimate of the elastic energy required to cross the barrier.\\
Measuring and discussing the stress-dependence of this local acceleration function $f$ in amorphous polymers from macroscopic creep experiments is the main focus of our article. Obviously, a macroscopic acceleration function can be determined from a simple comparison of the experimental linear and non-linear mechanical response of the polymer. However, the relationship between this macroscopic acceleration function  and the local one  is difficult to establish because of the development of stress heterogeneities. Recently, we have shown that during the stress relaxation resulting from a step-strain, the macroscopic acceleration function can be very different from the local one~\cite{belguise_weak_2021} and more precisely that it can exhibit a different scaling. However, during a step-strain the change in the macroscopic stress makes the estimate of the local acceleration challenging. As discussed by Belguise~\textit{et al.}~\cite{belguise_weak_2021}, it requires the analysis of experimental data in the light of numerical simulations accounting for material disorder.\\
During creep, the changes in the local stress distribution within the polymer are somehow different from step-strain because the macroscopic applied stress remains constant. As a consequence, the average local stress remains constant. So, we expect that the local and macroscopic acceleration functions remain roughly similar as long as the stress disorder is weak. In this paper, we study the apparition of non-linearities during creep experiments using an amorphous poly(etherimide) (PEI) polymer as a model system. We limit ourselves to the cases where the macroscopic field of deformation has a small amplitude (typically smaller than 15\%). In this strain range, we show that geometric non-linearities are negligible as compared to the one resulting from the stress-induced acceleration of configurational changes. This regime where a non linear mechanical response is observed in the absence of significant geometrical non linearities will be called subsequently the weak-non linear regime.\\
The paper is organized as follows: we first present the experimental protocol and the creep measurements carried out both in the linear and in the weakly non linear regimes. Then, we discuss the results and establish that the local acceleration function is $f=e^{-(\sigma/Y)^n}$ with $n=2 \pm 0.2$. From an estimate of the amplification of the stress applied locally on unrelaxed domains with slow relaxation times, we also discuss the effects of stress disorder.\\
%
%
\section{Experimental details}
\subsection{Poly(etherimide) polymer}
\begin{figure}
\centering
\includegraphics[width=1\linewidth]{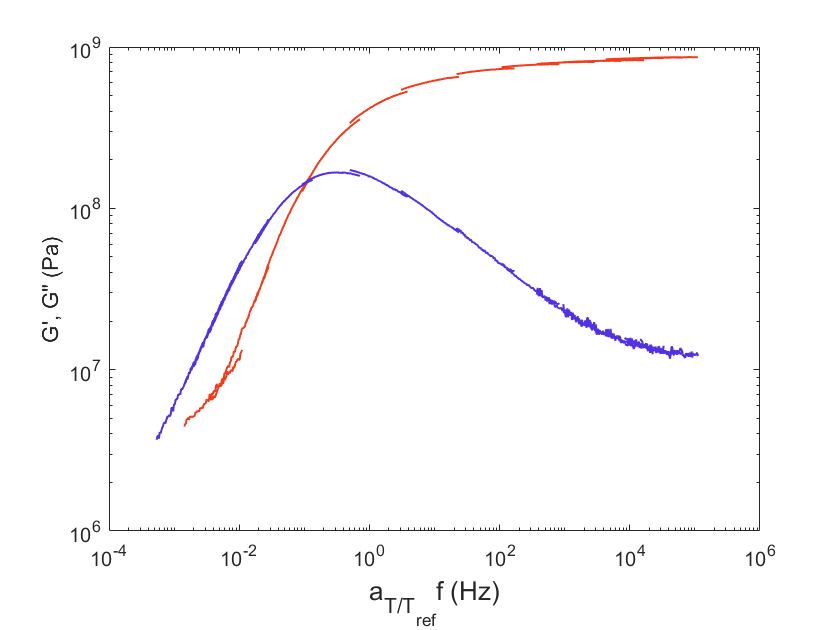}
\caption{Master curve giving the storage ($G'$, in red) and the loss ($G''$, in blue) shear moduli of the amorphous PEI in the linear range and at the reference temperature $T_{ref}=T_g=213\si{\celsius}$ (shear deformation $\gamma=0.05\%$).}
\label{fig:linear_master_curve_G}
\end{figure}

\begin{figure}
\centering
\includegraphics[width=1\linewidth]{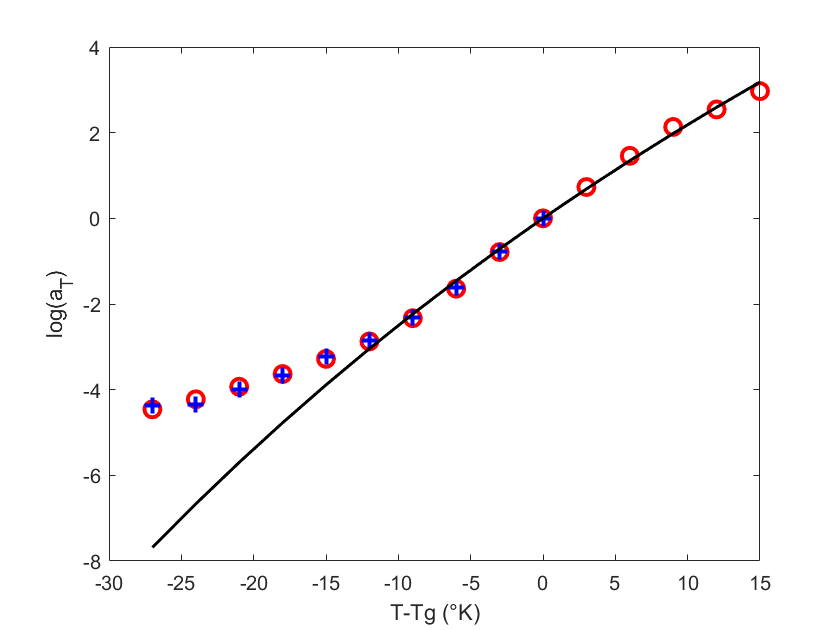}
\caption{Temperature-dependence of the logarithm of the shift factor $a_T$ determined from time-temperature superposition in the linear regime using either (\textcolor{red}{o}) $G'$ and $G"$ data or (\textcolor{blue}{+}) creep compliance data. $a_T$ values were measured directly from the frequency shifts (resp. the time shifts) of $G'$ and $G"$ (resp. $J(t)$) isotherms. The solid line is a fit of the data to William-Landel-Ferry (WLF) relationship within the temperature range $-10\si{\celsius}<T-T_g<15\si{\celsius}$ with $C_1 = 35$ and $C_2 = 150$~\si{\kelvin}.}
\label{fig:aT}
\end{figure}

This study has been carried out using injection molded poly(etherimide) (PEI) polymer sheets (Ultem 1010, Sabic) which were supplied by Arkema. The glass transition temperature of the used amorphous PEI grade is $T_g=213\si{\celsius}$ as measured by DSC at 10~\si{\celsius\per\minute} and its number molar mass is $M_n$=2~10$^4$~\si{\gram\per\mol}.\\
The linear shear viscoelastic properties of the PEI polymer were measured in a torsional mode using rectangular specimens 10x2x2 mm and a MCR 702 rheometer (Anton Paar, Austria). Fig.~\ref{fig:linear_master_curve_G} shows the resulting master curve obtained at $T_{ref}=213$~$\si{\celsius}$ and for a shear deformation $\gamma=0.05\%$. The corresponding shift factors $a_T$ are reported in Fig.~\ref{fig:aT} as red circles. As anticipated, the temperature-dependence of shift factors obeys William-Landel-Ferry (WLF) relationship within the temperature range $-10\si{\celsius}<T-T_g<15\si{\celsius}$.\\
%
%
\subsection{Creep measurements} 
Creep experiments in tensile mode were carried out using the same MCR702 rheometer as for the torsional measurements. Here, we focus on the creep behaviour in a weakly non-linear regime (applied stress less than 15~\si{\mega\pascal}, and strain less than $15~\%$) where geometrical non linearities can be neglected. In this regime, a relevant comparison between the linear and non linear creep responses requires that the compliance $J(t,\Sigma)=\epsilon(t)/\Sigma$ (where $\epsilon(t)$ is the measured strain and $\Sigma$ is the macroscopic applied stress) is determined within an accuracy less than a few percent. Such a requirement prevents any comparison based on experiments carried out using different samples. Indeed, their mechanical response can vary by as much as $10\%$ as a result of small uncertainties in their shape or of small misalignment's in the grips of the rheometer. In order to avoid such an experimental scatter, the following procedure was developed: (\textit{i}) the creep master curve of a given specimen is first determined in the linear regime from successive measurements at $\sigma=1$~$\si{\mega\pascal}$ and at increasing temperatures from 186 to 213~\si{\celsius} by 3~\si{\celsius} steps (we have checked experimentally that a 1~$\si{\mega\pascal}$ stress lies in the linear regime); (\textit{ii}) the non linear response of the \textit{same} specimen is subsequently measured at a given stress level and at increasing temperatures from 186 to 213~\si{\celsius} while \textit{it remains fixed within the grips of the rheometer}. This procedure is repeated with a different specimen for each of the considered non-linear stress levels in the range 3-15~$\si{\mega\pascal}$. 
Deformation up to 50\% can be reached, but in order to remain in the weak non linear regime, we limit ourselves to results for which the elongation is smaller than  15\% as mentioned above.\\
Indeed, such a procedure requires that the thermo-mechanical history of the specimen in between two successive creep sequences is erased. For that purpose, the specimen is annealed after each creep sequence under a zero applied stress at $T=T_g-3$~\si{\celsius} until it recovers its undeformed length within a 5\% accuracy. The corresponding annealing time varies from 40~$\si{\second}$ to 2000~$\si{\second}$ after creep measurements carried out from 186\si{\celsius} to 210\si{\celsius}.
After each isothermal strain recovery step, the specimen is subsequently allowed to cool down to the prescribed temperature of the next creep sequence in about 400~$\si{\second}$. Then, a 600~$\si{\second}$ isotherm is observed before the application of the next creep stress in order to achieve a uniform temperature within the sample.\\
Creep experiments were carried out using dog-bone specimens which were milled from 2~\si{\milli\meter} thick PEI plates. The center section of the dogbone was 1~\si{\milli\meter} wide and 10~\si{\milli\meter} long.
The elongation of the sample was measured with a µm accuracy under constant applied forces ranging from 2 to 30 \si{\newton}, \textit{i.e.} under a constant nominal stress (from 1 to 15 \si{\mega\pascal}). Any geometrical non linearity being neglected, the compliance is calculated using the nominal stress both in the linear and non linear regimes.\\
%
%
\section{Experimental results}
\subsection{Creep compliance}
\begin{figure} [!ht]
\centering
\includegraphics[width=1\linewidth]{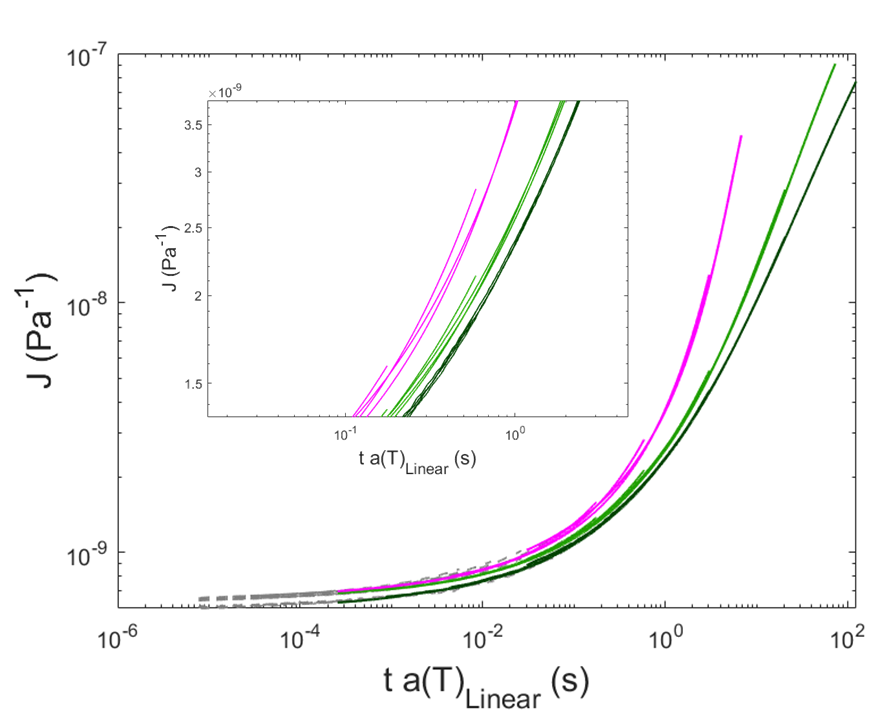}
\caption{Creep compliance, $J_L$, in the linear regime (for $\Sigma \leq 1~\si{\mega\pascal}$, in black) and $J_{NL}$, in the nonlinear regime, for $\Sigma=5$~\si{\mega\pascal} (green) and $\Sigma=10$~\si{\mega\pascal} (magenta) as a function of the reduced time $a_Tt$ at the reference temperature $T_{ref}=213\si{\celsius}$. The master curves in the non linear regime were established using the shift factors $a_T$ determined in the linear regime. Inset: details of the compliance curves showing imperfections in the time-temperature superposition in non-linear regime.}
\label{fig:J_lin_nonlin}
\end{figure}

As a reference, the master curve giving the creep compliance $J_L$ in the linear regime (black line in Fig.~\ref{fig:J_lin_nonlin}) was established at $T_{ref}=T_g=213\si{\celsius}$ from the superposition of the creep compliance isotherms at $\Sigma=1 \si{\mega\pascal}$. As shown in Fig.~\ref{fig:aT}, the corresponding shift factors $a_T$ are in perfect agreement with those deduced from the viscoelastic shear modulus measurements.
%
In addition to the linear compliance, the non-linear compliance $J_{NL}(\Sigma,a_Tt)$ is also reported in Fig.~\ref{fig:J_lin_nonlin} for $\Sigma=$5 and 10~\si{\mega\pascal}. These master curves were established using the shift factors $a_T$ which were determined in the linear regime. The main effect of applied stress in the non linear regime is to induce an overall increase in the compliance values that we interpret, as detailed below, as shift of the compliance to shorter times. The imperfect overlap of the isotherms (see inset in Fig.~\ref{fig:J_lin_nonlin}) indicates that the time-temperature superposition obeyed in the linear regime only poorly applies to the non-linear regime, which was noticed previously~\cite{oconnell_large_1997,oconnell_no_2002}. However, the amplitude of the resulting scatter is much smaller than the magnitude of the observed time shift with respect to the linear regime. As a consequence, we will neglect in what follows any effect of temperature in the description of the acceleration of the creep behavior under stress.\\
\subsection{Acceleration function}
\begin{figure} [!ht]
\centering
\includegraphics[width=1\linewidth]{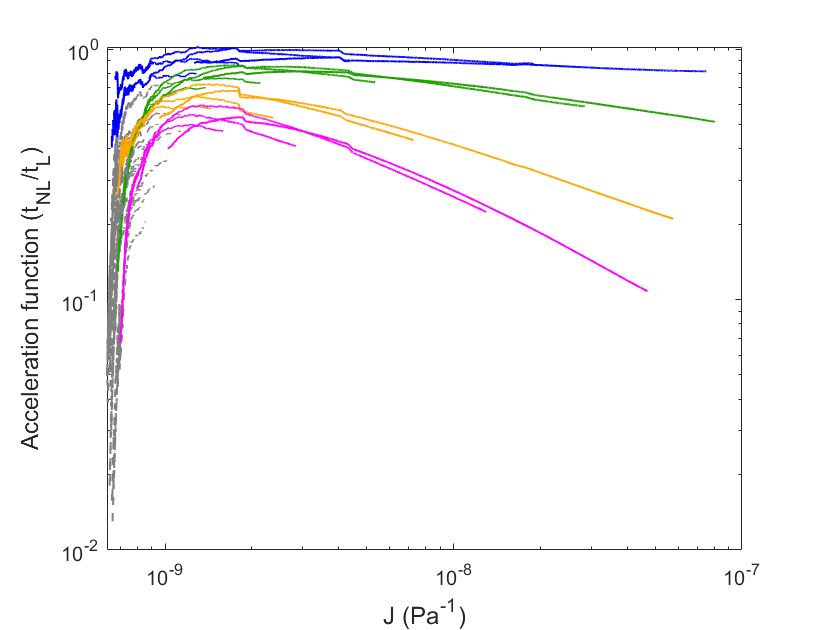}
\caption{Macroscopic acceleration function $F$ as a function of the compliance $J$ for increasing values of the applied stress $\Sigma$. $\Sigma=$~3~\si{\mega\pascal} (blue), 5~\si{\mega\pascal} (green), 8~\si{\mega\pascal} (orange) and 10~\si{\mega\pascal} (magenta).  When $J \lesssim 10^{-9}$~\si{\per\pascal} (dashed gray lines on the graph), $F$ is strongly influenced by the thermo-mechanical history of the specimen.}
\label{fig:F_macro}
\end{figure}

\begin{figure}
     \centering
     \begin{subfigure}[b]{1\linewidth}
         \centering
         \includegraphics[width=\textwidth]{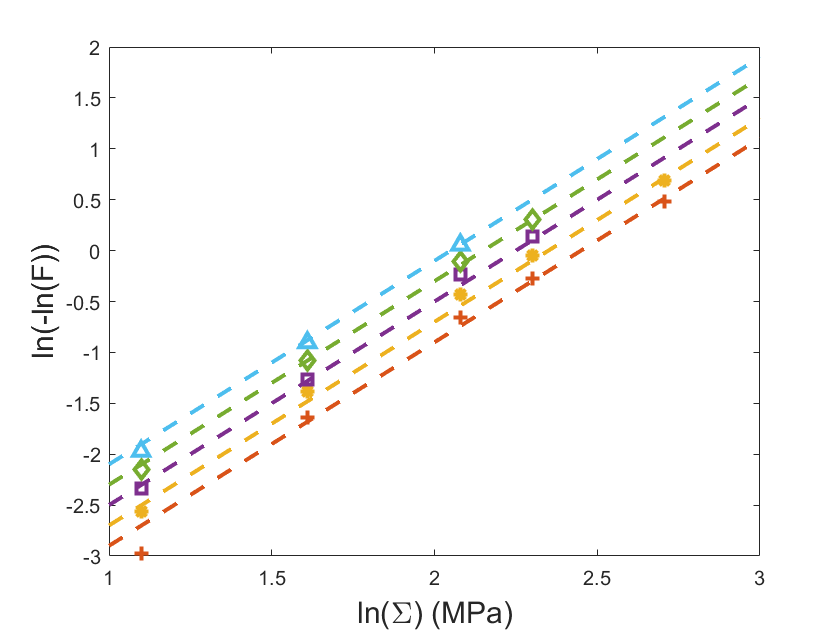}
         \caption{}
         \label{fig:pente_n}
     \end{subfigure}
     \hfill
     \begin{subfigure}[b]{1\linewidth}
         \centering
         \includegraphics[width=\textwidth]{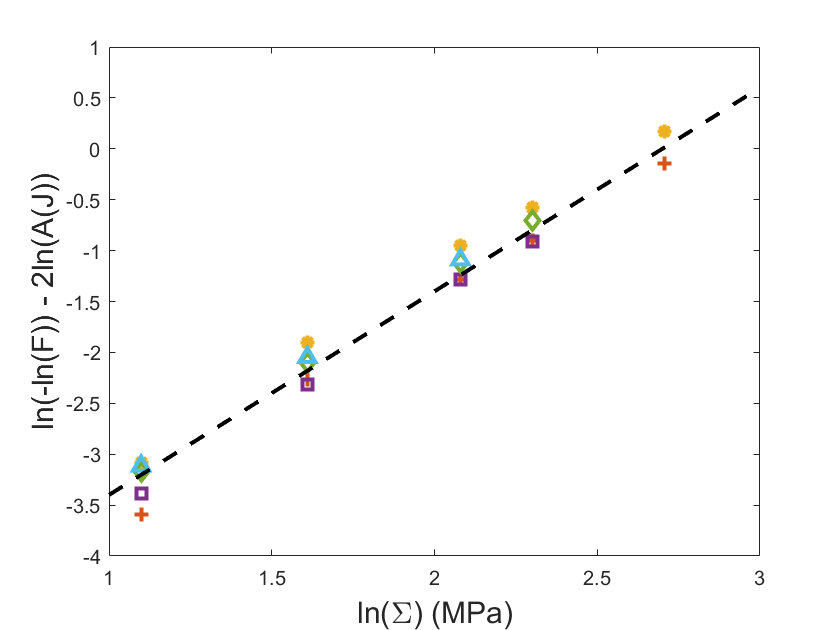}
         \caption{}
         \label{fig:pente_n_corr_A}
     \end{subfigure}
        \caption{Stress dependence of the macroscopic acceleration function $F$. (a) $\ln(-\ln(F))$ as a function of $\ln(\Sigma)$ for increasing values of the compliance $J$.  Dotted lines correspond to a slope $n=2$ fitted for each $J$ values. (b) $\ln(-\ln(F)) - 2\ln(A)$ as a function of $\ln(\Sigma)$, where $\ln(A)=\ln(\frac{\Sigma}{Y(J/J_0)})$. From bottom to top: (\textcolor{red}{+}) $J=$~3.04~10$^{-9}$, (\textcolor{yellow}{\textbullet})~4.60~10$^{-9}$, (\textcolor{purple}{$\square$})~6.98~10$^{-9}$ (\textcolor{green}{$\diamond$})~1.06~10$^{-8}$ and (\textcolor{blue}{$\triangle$})~1.61~10$^{-8}$~\si{\per\pascal}}
        \label{fig:three_graphs}
\end{figure}

\begin{figure}
     \centering
     \begin{subfigure}[b]{1\linewidth}
         \centering
         \includegraphics[width=\textwidth]{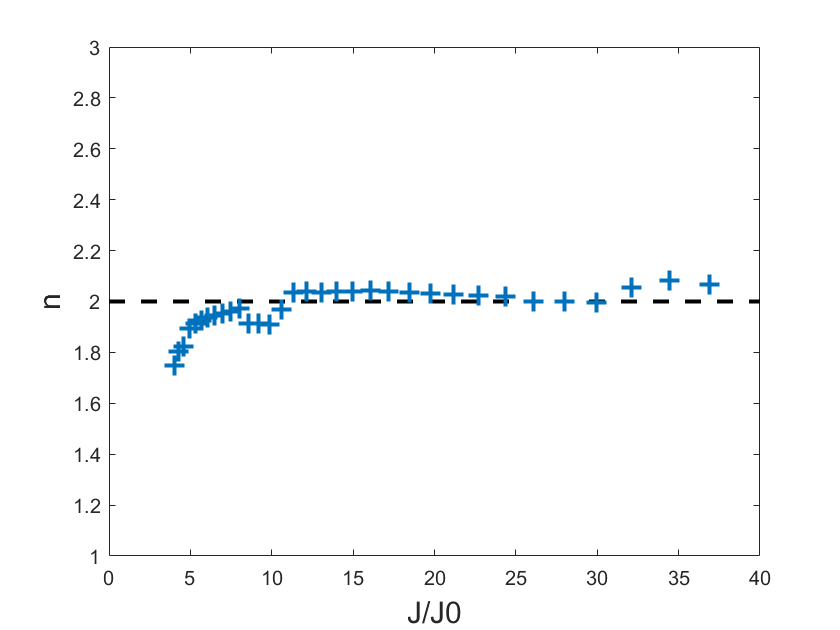}
         \caption{}
         \label{fig:nvJ}
     \end{subfigure}
     \hfill
     \begin{subfigure}[b]{1\linewidth}
         \centering
         \includegraphics[width=\textwidth]{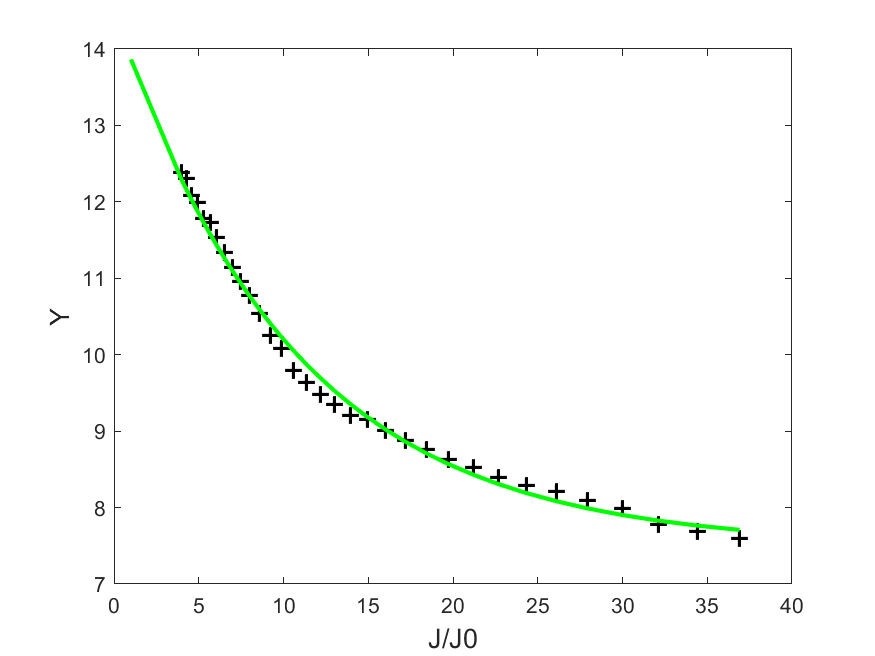}
         \caption{}
         \label{fig:YvJ}
     \end{subfigure}
        \caption{Values of the parameters of the fit of experimental $F(\Sigma,J)$ data to Eqn.~\ref{eq:F_1} as a function of the normalized compliance $J/J0$. (a) $n$ as a function of normalized compliance $J/J_0$. The dotted line corresponds to the theoretical value, $n=2$. (b) $Y$ as a function of normalized compliance $J/J_0$. The green line corresponds to fit of experimental data to an empirical decreasing exponential function, $Y=7e^{-\frac{J}{10.5 J0}} + 7.5$.} 
        \label{fig:Y_N}
\end{figure}

We now turn to a quantitative description of the time shift of the creep response in the non linear regime from the definition of an acceleration function. Following the generalised Voigt model formalism~\cite{ferry_viscoelastic_1980}, the creep function $J_L(t)$ of a material in the linear regime can always be written as
\begin{equation}
J_L(t)=\int_{-\infty}^{+\infty}L(\tau)\left(1-e^{-\frac{t}{\tau}}\right)dln(\tau)  ,  
\end{equation}
where $L(\tau)$ is the spectrum of relaxation times. Under the effect of stress in the non linear regime, the relaxation time of each element is assumed to be reduced by a function of the local stress, $f(\sigma)$. Here, we make the assumption that the local stress $\sigma$ is homogeneous and equal to the macroscopic one, $\Sigma$, so that the local acceleration $f(\sigma)$ function can be replaced by a macroscopic one, $F(\Sigma)$. Hence, the non linear creep $J_{NL}(\Sigma,t)$ writes
\begin{eqnarray}
J_{NL}(\Sigma,t)&=\int_{-\infty}^{+\infty}L(\tau)\left(1-e^{-\frac{t}{F(\Sigma)\tau}}\right)dln(\tau) \nonumber \\
&= J_L\left(\frac{t}{F(\Sigma)}\right) .
\end{eqnarray}   
 Thus, assuming that the stress is homogeneous in the sample leads to the simple result that the $J_{NL}(\Sigma,t)$  curves are shifted horizontally with respect to the linear one, $J_{L}(t)$, in a log scale by a coefficient that is constant for a given value of $\Sigma$. As it can be seen in Fig.~\ref{fig:J_lin_nonlin}, this does not correspond to our measurements which show that the horizontal shift between $J_{L}(t)$ and $J_{NL}(\Sigma,t)$  depends on the value of the compliance. However, we can still define a macroscopic acceleration function $F(\Sigma,J)$ such that $J_{NL} (t)=J_L(t/F(\Sigma,J))$. This function $F$ is plotted in Fig.~\ref{fig:F_macro} as a function of the compliance for various values of the applied stress. When $J \lesssim 10^{-9}$~\si{\per\pascal}, it turns out that the acceleration function exhibits very low values which are much scattered. As indicated by separate experiments where the thermo-mechanical history of the specimen was varied before creep (results not shown), this peculiar effect originates from the non-linear aging of the material. Indeed, measurements are performed slightly below the glass transition, at temperatures between 186~\si{\celsius} and 201~\si{\celsius}, where aging is significant at the time scale of the experiments. As extensively studied by Struik~\cite{struik_physical_1977} and many others, creep in this aging regime exhibits a time-dependence of the compliance which depends crucially on the thermal history of the sample. Such a behavior reflects the fact that material domains have been trapped in a state far from the thermodynamics equilibrium and evolve slowly towards it. As a consequence, the associated apparent time-temperature coefficients deviate from the classical WLF law~\cite{oconnell_arrhenius-type_1999}. A description of these aging effects on non linear creep is beyond the scope of this study and we will restrict in what follows to the description of the acceleration function for compliance greater than $10^{-9}$~\si{\per\pascal}. This range corresponds to compliance measurements carried out at temperatures greater than $T_g$-15~\si{\kelvin} for which we know from Fig.~\ref{fig:aT} that we are at thermodynamics equilibrium.\\
%
As shown in Fig.~\ref{fig:F_macro}, the acceleration function when $J\gtrsim 10^{-9}$~\si{\per\pascal} shows a plateau followed by a decrease with compliance whose magnitude is enhanced with the applied stress. Within experimental accuracy, we were not able to detect any significant effect of the temperature on $F$ for the considered stress and temperature ranges.\\
Taking inspiration from the theoretical expression derived by Long~\textit{et al.}~\cite{long_dynamics_2018,dequidt_heterogeneous_2016} for the local acceleration function $f$, we checked whether the experimental macroscopic acceleration function $F$ can be fitted using the following expression
\begin{equation}
 F=e^{-\left(\frac{\Sigma}{Y}\right)^n}.
 \label{eq:F_1}
\end{equation}
For that purpose, experimental values of $\ln(-\ln(F))$ were plotted in Fig.~\ref{fig:pente_n} as a function of $\ln(\Sigma)$ for increasing values of the compliance. Remarkably, a linear plot is obtained for all the $J$ values with an exponent $n=2 \pm 0.2$ which is independent on $J$ (see Fig.~\ref{fig:nvJ}) and very close to the theoretical value ($n=2$) derived by Long~\textit{et al}.
%

Conversely, the vertical shift of the $\ln(-\ln(F))$ vs $\ln(\Sigma)$ plots when $J$ increases indicates that the parameter $Y$ of the fit is a function of the compliance. This dependency is further considered in Fig.~\ref{fig:YvJ}, where $Y$ values deduced from the fit to Eqn.~\ref{eq:F_1} (with $n=2$) are plotted as a function of the normalized compliance $J/J_0$, where $J_0=J(t=0)$.
%
In what follows, we discuss the possible origin of the observed decrease in $Y$ at increasing $J/J_0$ from the development of stress inhomogeneities at the scale of nanometric domains.\\
\section{Discussion}
\subsection{Local stress and amplification factor}
In what follows, we will first assume that before relaxation - or plastic deformation or rearrangement- the response of the material is elastic and linear. Indeed, the macroscopic yield strain of amorphous polymers is about a few percent, thus suggesting that the the strain at which rearrangements are initiated is also about a few percent. At such a low strain levels, the elastic response is expected to be linear. This implies that the non-linear response originates only from the rearrangement of the local domains, as hypothesised in most of the models of amorphous plasticity~\cite{langer_shear-transformation-zone_2008,argon_theory_1973,dequidt_heterogeneous_2016}. In other words, we will make the assumption that the mechanical response of the system - before relaxation - remains intrinsically linear at the local scale and that the observed macroscopic non linear creep behaviour originates from the increasing number of relaxed domains. During creep, the stress field thus becomes more and more heterogeneous and the local stresses achieved within unrelaxed domains increase. Let $\phi$ be the volume fraction of these  relaxed domains. As the average or macroscopic stress $\Sigma$ remains constant during creep, the macroscopic strain will increase because the stress is carried by a decreasing fraction $1-\phi$ of unrelaxed domains. As a consequence, we assume that $J_{NL}(\Sigma,t)$ is a function of $\phi$ only.\\
At the scale of an individual unrelaxed domain $i$, the local stress writes $\sigma_i=A_i(\phi)\Sigma$, where $A_i$ is the stress amplification factor of the considered domain. As the fraction $\phi$ of relaxed domains determines the compliance, it should be kept in mind that $A_i$ is a increasing function of $J$. Based on these assumptions, we will develop below a simple mean field approximation for the acceleration function.\\
\subsection{Approximate mean field model}
\begin{figure} [!ht]
\centering
\includegraphics[width=1\linewidth]{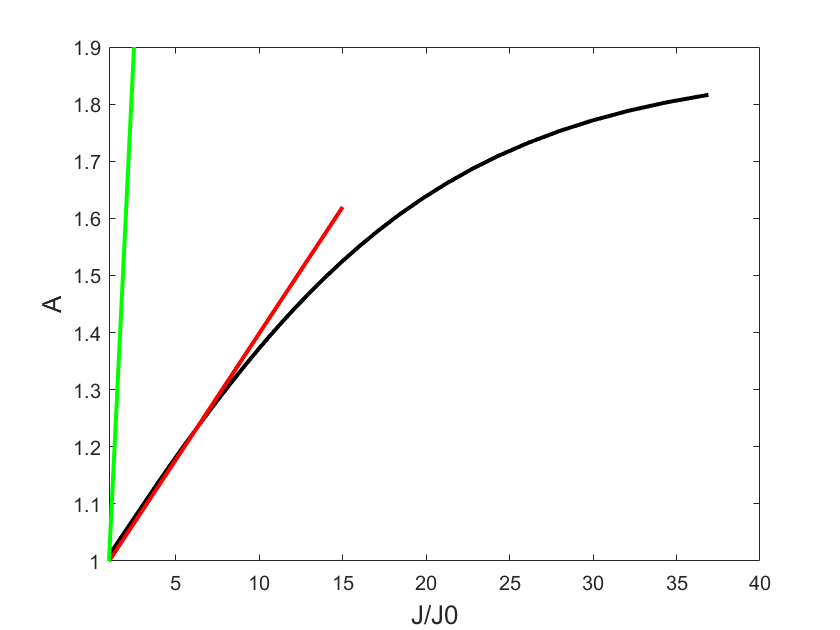}
\caption{Stress amplification factor $A(J)$ as a function of the normalized compliance $J/J_0$. When $J/J_0=1$, stress homogeneity within the unrelaxed sample implies that $A=1$. The green line corresponds to Eq.~\ref{eq:AJ}. The red line is a linear fit of experimental $A(J/J_0)$ data at low compliance.} 
\label{fig:AJ}
\end{figure}

\begin{figure} [!ht]
\centering
\includegraphics[width=1\linewidth]{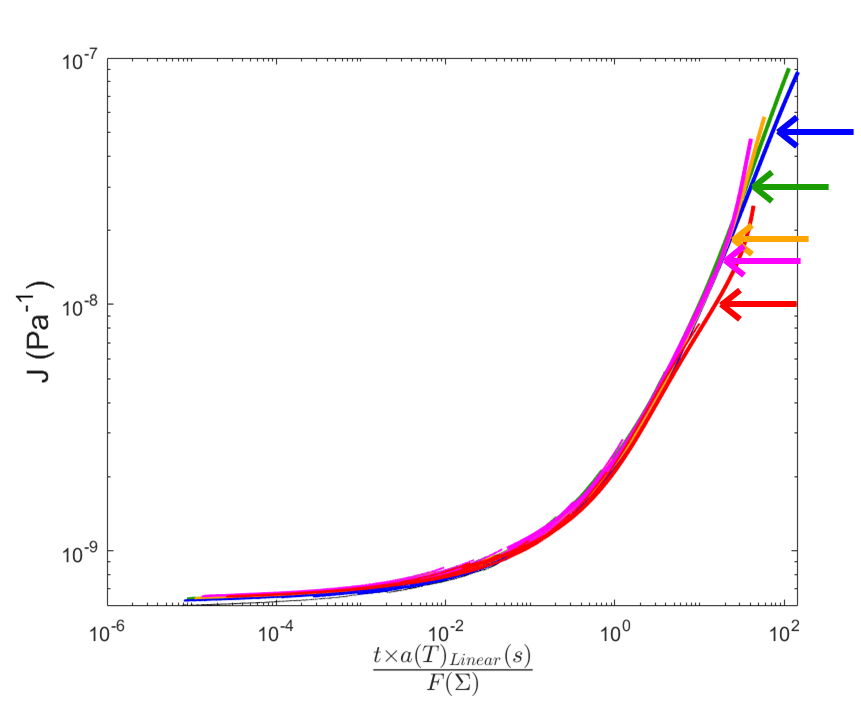}
\caption{Master curve giving the compliance $J$ as function of the reduced time $a_T t \exp^{-\left(\frac{A(J)\Sigma}{Y_0}\right)^n}$. Applied stress $\Sigma$=3~\si{\mega\pascal} (blue), 5~\si{\mega\pascal} (green), 8~\si{\mega\pascal} (orange)  10~\si{\mega\pascal} (magenta) and 15~\si{\mega\pascal} (red). The arrows with the corresponding colors indicate the 15\% strain limit used when processing the data.} 
\label{fig:Jmaster}
\end{figure}

It is out of the scope of this paper to solve completely the theoretical problem of the sequence of domains' relaxation and of the associated stress field redistribution. However, we develop in what follows a very crude description of the behavior of the polymer based on a mean field approximation. We will assume that all the relaxed domains do no longer sustain any stress and behave like cavity while the all non-relaxed domains bear the same stress $\sigma$. In order to estimate this stress, we use the Palierne's self-consistent model~\cite{palierne_linear_1990} for the modulus of systems with inclusions. Let $G_0$ be the glassy modulus of the individual unrelaxed domains. The macroscopic modulus derived from Palierne's model for our system with cavities within a matrix of unrelaxed domains writes (Eqn.~4.2 in reference~\cite{palierne_linear_1990}) 
\begin{equation}
 G(t)=G_0 \frac{1-\phi}{1+2/3\phi}.   
\end{equation}
The compliance can thus be expressed as
\begin{equation}
J(t)=J_0\frac{1+2/3\phi(t)}{1-\phi(t)}.
    \label{eq:J_phi}
 \end{equation}
Meanwhile, the local stress $\sigma$ is sustained by unrelaxed zones only. Within the framework of the assumption that all the non-relaxed domains are identical mechanically, we have therefore $\sigma (1-\phi)=\Sigma$, as the macroscopic stress is an average of the local one over the unrelaxed domains. Hence, the stress amplification factor writes
\begin{equation}
    A=\frac{\sigma}{\Sigma}=\frac{3}{5}J(t)/J_0+\frac{2}{3}.
    \label{eq:AJ}
\end{equation}
According to this very crude approximation, the stress amplification factor $A$ is thus assumed to increase linearly with the compliance.\\
An experimental estimate of the dependence of the amplification factor on compliance can be obtained from the measured $Y(J/J_0)$ relationship shown in Fig.~\ref{fig:YvJ} if one assumes that the observed decrease in $Y$ with increasing $J$ is only due to stress amplification. Assuming that the microscopic acceleration is given by $f=e^{-(\sigma/Y)^2}$, the macroscopic stress acceleration function $F$ can be written as follows in the crude approximation that the microscopic stress distribution in the unrelaxed domains can be replaced by a single value $A(J)\Sigma$
\begin{equation}
    F=e^{-\left(\frac{A(J)\Sigma}{Y_0}\right)^2},
    \label{eq:F_2}
\end{equation}
where $Y_0$ is the value of $Y$ when $J/J0=1$. According to Eq.~\ref{eq:F_1} and \ref{eq:F_2}, the stress amplification factor can thus be expressed as $A(J)=Y_0/Y(J)$. Accordingly, $A=1$ when $J/J_0=1$. From a fit of the experimental $Y(J/J_0)$ data in Fig.~\ref{fig:YvJ} to a decreasing exponential function, the value of $Y_0$ was extrapolated to 14~\si{\mega\pascal}. Using this value, $A(J)$ data were calculated from the fitted $Y(J)$ values. As expected, a knowledge of the stress amplification factor allows to shift all the macroscopic acceleration function data to a single master curve as shown in Fig.~\ref{fig:pente_n_corr_A}.\\
In Fig.~\ref{fig:AJ}, the stress amplication factor $A$ is reported as a function of the normalized compliance $J/J_0$. According to the theoretical mean field model, $A$ is linearly increasing with $J$ at low compliance. However, the corresponding slope (red line in Fig.~\ref{fig:AJ}) is smaller than that predicted by the mean field model (the green line in Fig.~\ref{fig:AJ}) by a factor of more than $10$. This discrepancy is likely to be due to our crude mean field approximation where we have assumed an homogeneous stress for all the non-relaxed domains and a vanishing stress for unrelaxed ones while a more complex stress distribution should be achieved in the real system. More precisely, our mean field approach probably discards many specific features of stress localization which result from the mechanical coupling between domains. Indeed, Finite Element simulations of the mechanics of amorphous polymers using a more refined stochastic continuum mechanics description of the disorder show that stress amplification occurs within band oriented at 45~deg. with respect to the applied stress~\cite{masurel_role_2015}. The development of such bands evidences collective relaxation mechanisms which are not accounted for in our mean field approximation. Nevertheless, the later allows to capture the basic ingredients of the relationship between compliance and strain amplification.\\
%
Finally, we have shown that it is possible to determine from weakly non linear creep experiments a macroscopic acceleration law (Eq.~\ref{eq:F_1}) with an exponent $n$ close to $2$, in agreement with the the theoretical value predicted by Long~\textit{et al.}~\cite{long_dynamics_2018}. However, the local stress is progressively amplified over unrelaxed domains as creep proceeds. The fact that the local stress is different from the macroscopic one has already been noticed for step strain experiments~\cite{belguise_weak_2021} but in creep the macroscopic stress is constant in time, thus allowing for a direct measurement of the exponent of the acceleration function - as opposed to the case of a step strain where the stress is decreasing with time. We also observed that the parameter $Y$ of the macroscopic acceleration function decreases with increasing compliance which suggests that the stress field becomes more and more disordered with time under creep.\\
Lastly, our approach allows to derive the following rescaling for compliance in the weakly non-linear creep regime
\begin{equation}
    J_{NL}(\Sigma,t)=J_L \left( t e^{-\left(\frac{A(J)\Sigma}{Y_0}\right)^2} \right),
    \label{eq:Jfin}
\end{equation}
where $A(J)=Y_0/Y(J)$. Using the above expression and the fitted values of $Y(J)$ (green line in Fig.~\ref{fig:YvJ}), it was possible to collapse with great accuracy all the non-linear creep data on a single master curve as shown in figure~\ref{fig:Jmaster}. We thus confirm that the local acceleration function is similar to the macroscopic one, up to a prefactor that depends only on the amplitude of the compliance $J$.\\
In this figure, we have also indicated by arrows the compliance values at which the strain reaches the limiting value used for data processing, \textit{i.e.} 15\%. It emerges that this strain limit corresponds roughly to the reduced time above which the rescaled compliance curves do no longer merge on the master curve, probably as a result of the development of geometrical non linearities which are beyond the scope of this work. 
%
%
\section{Conclusion}
In this paper we have investigated the weak nonlinear mechanical response of an amorphous polymer (PEI) in the vicinity of its glass transition. The non linear creep behavior of the polymer in this regime was interpreted as a consequence of the acceleration of the configurational rearrangements of local domains under the action of stress. This acceleration was accounted for by a shift of the relaxation times of the domains by a factor which depends exponentially on the square of the local stress, according to the theoretical law derived by Long~\textit{et al.}~\cite{long_dynamics_2018}. Moreover, we have observed that the local stress sustained by unrelaxed domains that bear the macroscopic stress increases with the compliance as a result of the enhanced disorder. From macroscopic creep compliance data, we have been able to measure the average amplification factor of the local stress versus the macroscopic one. This amplification factor modifies only the prefactor of the stress in the acceleration function, but not its scaling.\\
From a crude mean field approximation, we are able to capture qualitatively the linear dependence of the stress amplification factor during the initial stages of creep. A more refined mean field description of the weakly non linear response of the polymer would require that the stress-induced shift in the relaxation time distribution is accounted for as creep proceeds, together with local stress distribution. Such a model will be the topic of a forthcoming paper.\\
\section{Dedication}
This article is dedicated to Helene Montes, our colleague and friend, who combined great scientific qualities with profound humanity, and who recently passed away while we were working on this project.\\
%
\begin{acknowledgments}
We wish to acknowledge Jean-Claude Mancer for his kind help in the realization of the creep specimens.
\end{acknowledgments}
%
%
\bibliography{weak_non_lin_creep_biblio}
\end{document}